# A Non-Volatile All-Spin Non-Binary Matrix Multiplier: An Efficient Hardware Accelerator for Machine Learning

Rahnuma Rahman and Supriyo Bandyopadhyay, *Fellow, IEEE*

*Abstract--* **We propose and analyze a compact and *non-volatile* nanomagnetic (all-spin) non-binary matrix multiplier performing the multiply-and-accumulate (MAC) operation using two magnetic tunnel junctions – one activated by strain to act as the multiplier, and the other activated by spin-orbit torque pulses to act as a domain wall synapse that performs the operation of the accumulator. It has two advantages over the usual crossbar-based electronic non-binary matrix multiplier. First, while the crossbar architecture requires $N^3$ devices to multiply two $N{\times}N$ matrices, we require only $2N^2$ devices. Second, our matrix multiplier is non-volatile and retains the information about the product matrix after being powered off. Here, we present an example where each MAC operation can be performed in ~5 ns and the maximum energy dissipated per operation is ~$60N_{max}$ aJ, where $N_{max}$ is the largest matrix size. This provides a very useful hardware accelerator for machine learning and artificial intelligence tasks which involve the multiplication of large matrices. The non-volatility allows the matrix multiplier to be embedded in powerful non-von-Neumann architectures, including processor-in-memory. It also allows much of the computing to be done at the edge (of internet-of-things) while reducing the need to access the cloud, thereby making artificial intelligence more resilient against cyberattacks.**

## I. Introduction

Artificial intelligence (AI) is pervasive and ubiquitous in modern life (smart cities, smart appliances, autonomous self-driving vehicles, information processing, speech recognition, patient monitoring, etc.). Estimates by OpenAi predict an explosive growth of computational requirements in AI by a factor of $100\times$ every two years, which is a $50\times$ faster rate than Moore's law governing the evolution of the chip industry [1]. Most AI applications leverage machine learning (or deep learning based on neural networks) to perform two primary functions – training and inference. Algorithms for these tasks require multiplication of large matrices, such as in updating the synaptic weight matrices in deep learning networks (which is an essential feature of training a neuronal circuit), solving combinatorial optimization problems with Ising machines, etc. A deep neural network (DNN) is essentially a sequence of layers, each connected to the next through a matrix multiplication $[x] \to [M][x]$ representing synaptic connections. The input to the ($m$+1)-th layer is related to the $m$-th layer as $x_i^{m+1} = f\left( \sum_j M_{ij}^m x_j^m \right)$, where $f$ is a non-linear activation function. Hardware accelerators that can perform matrix multiplications rapidly and efficiently are therefore very attractive since they can speed up AI tasks immensely. They are particularly useful in computer vision [2], image and other classification tasks [3], approximate computing [4], speech recognition [5], patient monitoring [6] and biomedicine [7].

The earliest ideas for devising hardware-based matrix multipliers date back to 1909. Percy Ludgate conceived of a machine made of mechanical parts that was understandably unwieldy, slow and unreliable [8]. Modern matrix multipliers employ electronic charge-based circuitry that are fast, convenient and reliable [9], but also energy-hungry and volatile, i.e. they lose all information once powered off. Recently, matrix multipliers have been implemented with optical networks [10, 11], which can be energy-efficient and fast, but their drawback is the large footprint. They too are usually volatile since they typically use capacitors for the accumulation operation in matrix multiplication. In this paper, we present an all-magnetic (all-spin) implementation of a matrix multiplier, which is energy efficient, fast and has a much smaller footprint than its optical or electronic counterparts. Its most important advantage is that it is truly *non-volatile* and hence the matrix products can be stored indefinitely in the device after powering off.

---

This work was supported by the US National Science Foundation under grants CCF-2001255 and CCF-2006843.

The authors are with the Department of Electrical and Computer Engineering, Virginia Commonwealth University, Richmond, VA 23284, USA (email: rahmanr3@vcu.edu, sbandy@vcu.edu )

A slightly shorter version of this article has been accepted for publication by the IEEE Transactions on Electron Devices. DOI: 10.1109/TED.2022.3214167.

Consider the matrix multiplication operation $c_{ij} = \sum_m a_{im} b_{mj}$. This operation consists of multiplying pairs of numbers (one member of the pair picked from a row of the multiplier matrix and the other from a column of the multiplicand matrix) and then adding up the products of the pairs to produce an element of the product matrix. Thus, one would need: (1) a "multiplier" to multiply pairs of numbers, and (2) an "accumulator" which accumulates the individual products and adds them up. These are the two ingredients of a hardware accelerator for matrix multiplication. In this work, we implement the multiplier with a single straintronic magnetic tunnel junction (MTJ) and the accumulator with another magnetic tunnel junction (driven by spin-orbit torque) acting as a domain wall synapse [12]. In the next two sections, we describe the multiplier and the accumulator.

## II. MULTIPLIER

A schematic of the proposed multiplier is shown in Fig. 1. It consists of a straintronic elliptical MTJ that has a (magnetically) "hard" layer and a "soft" layer, separated by an intervening insulating spacer layer. Any residual dipole interaction between the hard and the soft ferromagnetic layers creates an effective magnetic field $H_d$ in the soft layer that is directed along the latter's major axis (easy axis) in a direction opposite to the magnetization of the hard layer. The soft layer is magnetostrictive and placed in elastic contact with an underlying *poled* piezoelectric thin film deposited on a conducting substrate (this construct constitutes a 2-phase multiferroic). Two electrically shorted electrodes, delineated on the piezoelectric film, flank the MTJ, while the back of the substrate is connected to ground.

When a (gate) voltage $V_G$ is applied to the shorted electrode pair, it generates biaxial strain in the piezoelectric film pinched between the two electrodes, which is transferred to the elliptical soft layer. The strain is compressive along the major axis and tensile along the minor axis of the soft layer, or vice versa, depending on the voltage polarity [13]. With the right voltage polarity, the biaxial strain will rotate the soft layer's magnetization away from the major axis of the ellipse (the magnetic easy axis) towards the minor axis (magnetic hard axis) because of the Villari effect. The rotation is opposed by the magnetic field $H_d$ which would like to keep the magnetization pointing along the initial orientation along the major axis. The interplay of these two effects ultimately makes the magnetization settle into an orientation that subtends some angle $\theta_{ss}$ with the major axis (or the magnetization of the hard layer). The value of $\theta_{ss}$ depends on the applied strain (or gate voltage $V_G$) and $H_d$. It corresponds to the location of the potential energy minimum in the presence of both strain and $H_d$. We show later that the potential energy has a deep minimum (many times the thermal energy $kT$) at $\theta_{ss}$, which makes $\theta_{ss}$ very stable against thermal noise. Because the hard layer's magnetization remains unaffected by strain, $H_d$ does not change with $V_G$. Therefore, as we change $V_G$ and the resulting strain, we will change $\theta_{ss}$ and consequently the MTJ resistance which depends on $\theta_{ss}$. A basic straintronic MTJ (s-MTJ), whose resistance was modulated with gate-induced strain, was experimentally demonstrated in [14].

To implement the multiplier, a constant current source $I_{bias}$ is connected between the hard and soft layers of the s-MTJ (terminals '1' and '2'), as shown in Fig. 1(a). This drives a current through the s-MTJ. The gate voltage $V_G$ is applied at terminal '3' to generate the strain in the soft layer, and a fourth terminal is connected to the hard layer (common with terminal '1'), which outputs a voltage $V_0$. Terminal 2, connected to the soft layer, is grounded and hence $V_0 = R_{s-MTJ} I_{bias}$, where $R_{s-MTJ}$ is the resistance of the s-MTJ that can be altered by the gate voltage $V_G$ generating strain, as explained before.

*A. Rotation of the soft layer's magnetization due to the gate voltage*

We have modeled the rotation of the soft layer's magnetization as a function of the gate voltage $V_G$ in the presence of $H_d$ and thermal noise using stochastic Landau-Lifshitz-Gilbert simulations [15]. This allows us to find the $\theta_{ss}$ versus $V_G$ relation. The s-MTJ resistance is given by $R_{s-MTJ} = R_P + \frac{R_{AP} - R_P}{2}[1 - \cos\theta_{ss}]$, where, $R_P$ is the s-MTJ resistance when the magnetizations of the hard and soft layers are mutually parallel and $R_{AP}$ is the s-MTJ resistance when the magnetizations are antiparallel. From the $\theta_{ss}$ versus $V_G$ relation, we can therefore calculate the $1/R_{s-MTJ}$ (= $G_{s-MTJ}$) versus $V_G$ characteristic, which we show qualitatively in Fig. 1(b). With proper choice of the s-MTJ parameters, we can produce a *linear* region in the $G_{s-MTJ}$ vs. $V_G$ characteristic where $1/R_{s-MTJ} = 1/R_{AP} + \kappa(V_G - \delta) \Rightarrow G_{s-MTJ} = G_{AP} + \kappa(V_G - \delta)$ [$\kappa$ and $\delta$ are constants]. We show analytically in the Appendix that such a linear region exists and follows the above relation.

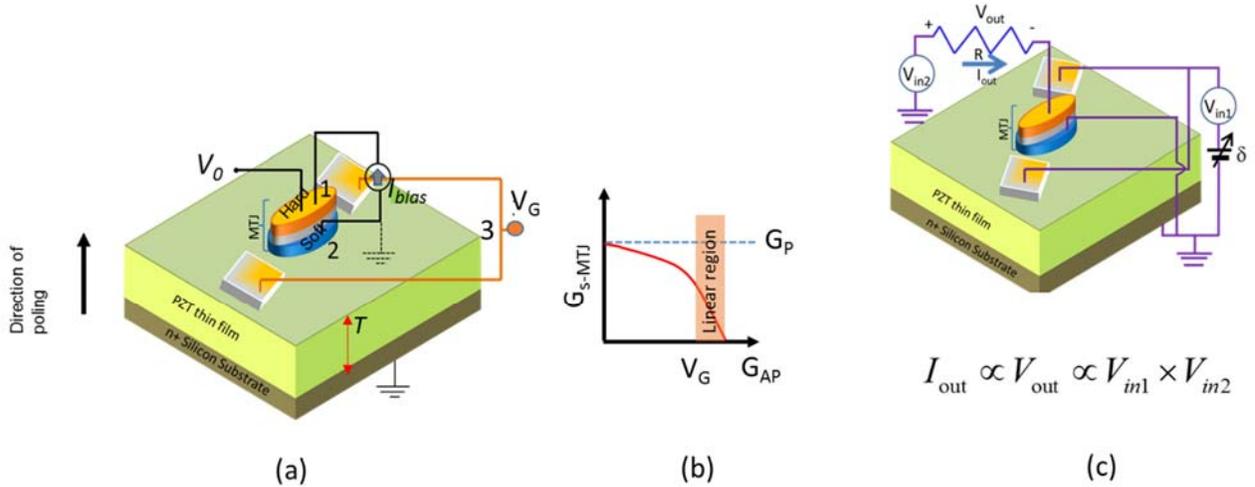

Fig. 1: (a) A straintronic magnetic tunnel junction (s-MTJ) configured to produce a linear region in the transfer characteristic $G_{MTJ}$ (magnetic tunnel junction conductance) versus $V_G$ (gate voltage). (b) The transfer characteristic showing the linear region. (c) An analog multiplier implemented with a single s-MTJ. The two operands are encoded in $V_{in1}$ and $V_{in2}$ and the product of them is encoded in $V_{out}$ or $I_{out}$. The s-MTJ is biased in the linear region of the transfer characteristic where the s-MTJ conductance is proportional to $(V_G - \delta)$ with $\delta$ being a bias voltage.

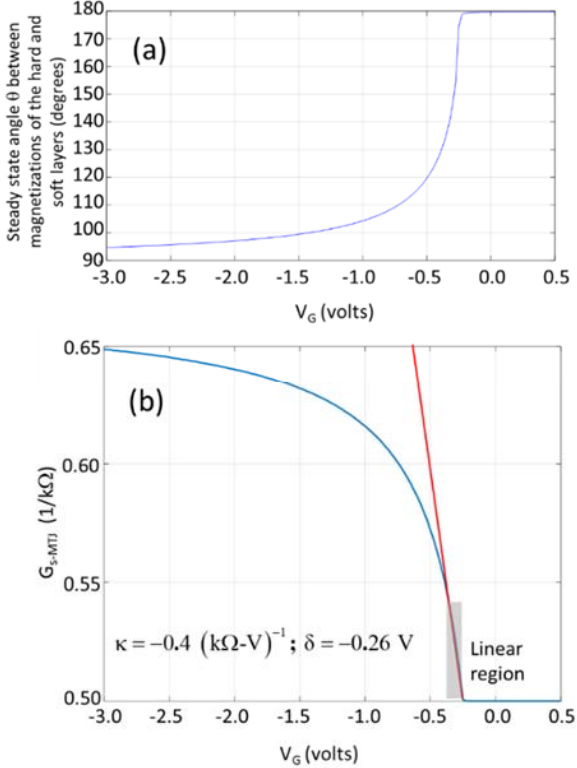

Fig. 2: Plots of (a) the steady-state value of the angle $\theta$ between the magnetizations of the hard and soft layers of the MTJ as a function of the gate voltage $V_G$ obtained from the stochastic Landau-Lifshitz-Gilbert simulation at room temperature (300 K). Because of thermal noise, which introduces randomness in the magnetization trajectory, this curve was obtained by averaging over 100 trajectories. (b) The $1/R_{MTJ}$ versus $V_G$ characteristic showing that there is a region (shaded in the figure) where the relation $G_{s-MTJ} = G_{AP} + \kappa(V_G - \delta)$ holds approximately. For this plot, we assumed $R_P$ = 1 k$\Omega$ and $R_{AP}$ = 2 k$\Omega$. The voltage $\delta$ and the constant $\kappa$ obtained by fitting a straight line to this plot are shown in the figure. We get $\kappa$ = -0.4 $\pm$ 0.045 (k$\Omega$-V)$^{-1}$ and $\delta$ = -0.26 $\pm$ 0.013 V. The various material parameters used to obtain these plots are given in Table I.

In Fig. 2, we plot the $\theta_{ss}$ versus $V_G$ characteristics obtained from the stochastic Landau Lifshitz Gilbert simulation and the resulting $G_{s-MTJ}$ versus $V_G$ plot. The simulation procedure is described in ref. [15] and the Appendix. The parameters for the elliptical soft layer of the s-MTJ used in the simulation are given in Table I. The soft layer is assumed to be made of Terfenol-D, which has large magnetostriction. The value of $H_d$ can be altered arbitrarily by applying an external magnetic field aligned with the dipole coupling field. The piezoelectric film is assumed to be (001) PMN-PT which has a large piezoelectric coefficient. The plot in Fig. 2(b) shows that there is indeed a region of $V_G$ where the MTJ conductance varies linearly with gate voltage and obeys the relation given above.

When the gate voltage $V_G$ is chosen to be in that linear region, the current $I_{out}$ in the construct of Fig. 1(c) becomes proportional to the product of two input voltages $V_{in1}$ and $V_{in2}$. These input voltages encode two elements of the multiplier and multiplicand matrices. Thus the device is Fig. 1(c) implements a multiplier. We show this explicitly in the next subsection.

**Table I: Parameters for the soft layer of the MTJ**

| Major axis dimension (L) | 800 nm |
|---|---|
| Minor axis dimension (W) | 700 nm |
| Thickness (d) | 2.2 nm |
| Saturation magnetization ($M_s$) | 8.5 $\times$ 10$^5$ A/m |
| Dipole coupling field ($H_d$) | 1000 Oe |
| Gilbert damping constant ($\alpha$) | 0.1 |
| Saturation magnetostriction ($\lambda_s$) | 600 ppm |
| Young's modulus | 120 GPa |
| Piezoelectric coefficient ($d_{33}$) | 1.5 $\times$ 10$^{-9}$ C/N |
| Piezoelectric layer thickness | 1 $\mu$m |

*B. Operation of the multiplier*

To understand how the multiplier works, refer to Fig. 1(c) and note that $V_{in1} = V_G - \delta$. Now, if $V_G$ is within the linear region in Fig. 2(b), then $G_{s-MTJ} = G_{AP} + \kappa(V_G - \delta) = G_{AP} + \kappa V_{in1}$. Also, note that the voltage dropped over the series resistor $R$ is

$$V_{out} = I_{out}R = \frac{R}{R + R_{s-MTJ}}V_{in2} \approx \frac{R}{R_{s-MTJ}}V_{in2} \approx RG_{s-MTJ}V_{in2} \quad (1)$$

if $R \ll R_{s-MTJ}$

Replacing $G_{s-MTJ}$ in Equation (1) with $G_{AP} + \kappa V_{in1}$, we get

$$V_{out} = RG_{AP}V_{in2} + R\kappa(V_{in1} \times V_{in2}) \approx R\kappa(V_{in1} \times V_{in2}) \text{ and}$$
$$I_{out} \approx \kappa(V_{in1} \times V_{in2}) \quad \text{since } R \ll R_{AP} \quad (2)$$

That implements a "multiplier" since the current $I_{out}$ flowing through the s-MTJ (which is also the current through the series resistor $R$) is proportional to the *product* of the two input voltages $V_{in1}$ and $V_{in2}$. The voltage $V_{out}$ is proportional to this current and hence it too is proportional to the product $V_{in1} \times V_{in2}$. Similar ideas were used to design probability composer circuits for Bayesian inference engines in the past [16]. In our case, $V_{in1}$ and $V_{in2}$ are voltage "pulses" of fixed width and varying amplitude. Their amplitudes are proportional to the two quantities to be multiplied.

Note from Fig. 2(b) that the linear region in the plot extends over a small voltage range of ~100 mV. Therefore, *for*

*this choice of parameters*, the amplitude of the $V_{in1}$ pulse should be no more than ~50 mV. Since we would like the two voltage pulses $V_{in1}$ and $V_{in2}$ to have similar limits on the amplitude, both should have an amplitude no more than 50 mV. We can, of course, increase the voltage range by redesigning with different parameters, but that will increase the energy dissipation per MAC operation.

## III. ACCUMULATOR

Next, imagine that the resistor $R$ of Fig. 1(c) is a heavy metal (HM) strip, on top of which we place a p-MTJ (which is an MTJ whose ferromagnetic layers have perpendicular magnetic anisotropy) with the soft layer in contact with the HM strip. We can insert a thin insulating layer and a thin metallic layer between the soft layer and the heavy metal, which will not impede the operation of the accumulator. This configuration is shown in Fig. 3(a). The current pulses $I_{out}$ coming out of the multiplier pass through the HM strip and because of spin-orbit interaction in that strip, they inject spins into the soft layer of the p-MTJ (through the thin insulating and metallic layers) during every pulse duration. That causes domain wall motion in the latter during each pulse owing to the spin Hall effect [17-19]. The distance a domain wall moves over the duration of a pulse is approximately proportional to the amplitude of the pulse and we show this later from micromagnetic simulations in the Appendix. The arrangement is shown in Fig. 3(b).

After a number of pulses, a fraction of the soft layer will have its magnetization parallel to that of the hard layer, a small fraction will be un-magnetized and will be the "domain wall" separating two domains, and the remainder of the soft layer will have its magnetization antiparallel to that of the hard layer. The fractions with parallel and anti-parallel magnetizations change with successive current pulses. This is the well-known basis of a domain wall synapse [12]. Here, we have used a p-MTJ in the spirit of ref. [12], but there is no reason why an MTJ with in-plane magnetic anisotropy cannot be used instead.

The conductance of the p-MTJ (measured between its hard and soft layers) is the conductance of the parallel combination of *three* conductors. The first is due to the segment where the soft layer's magnetization is parallel to that of the hard layer, the second is associated with the domain wall (DW), and the third is due to the segment where the hard and soft layers' magnetizations are mutually antiparallel [12], as shown in Fig. 3(c). If the domain wall in the soft layer of the p-MTJ is located at a distance $x$ from one edge and $L$ is the length of the soft layer, then [12]

$$G_{\text{p-MTJ}}(x) = G_{AP}\frac{x}{L} + G_{DW}\frac{w}{L} + G_P\left(\frac{L-x-w}{L}\right)$$
$$= \underbrace{G_{DW}\frac{w}{L} + G_P\left(1-\frac{w}{L}\right)}_{\text{constant A}} - \underbrace{\frac{(G_P - G_{AP})}{L}}_{\text{constant B}}x , \quad (3)$$

where $w$ is the domain wall width, $G_P$ is the p-MTJ conductance in the parallel state, $G_{AP}$ is the conductance in the antiparallel state and $G_{DW}$ is the conductance associated with the domain wall in the soft layer.

### A. Operation of the accumulator

To understand how the accumulator works, consider the fact that the quantity $a_{im}$ is encoded in the amplitude of the $m$-th voltage pulse of $V_{in1}$ and $b_{mj}$ is encoded in the amplitude of the $m$-th voltage pulse of $V_{in2}$. The $m$-th current pulse $I_{out}^m$ coming out of the multiplier and flowing through the HM strip therefore has an amplitude $I_{out}^m \propto a_{im} \times b_{mj}$.

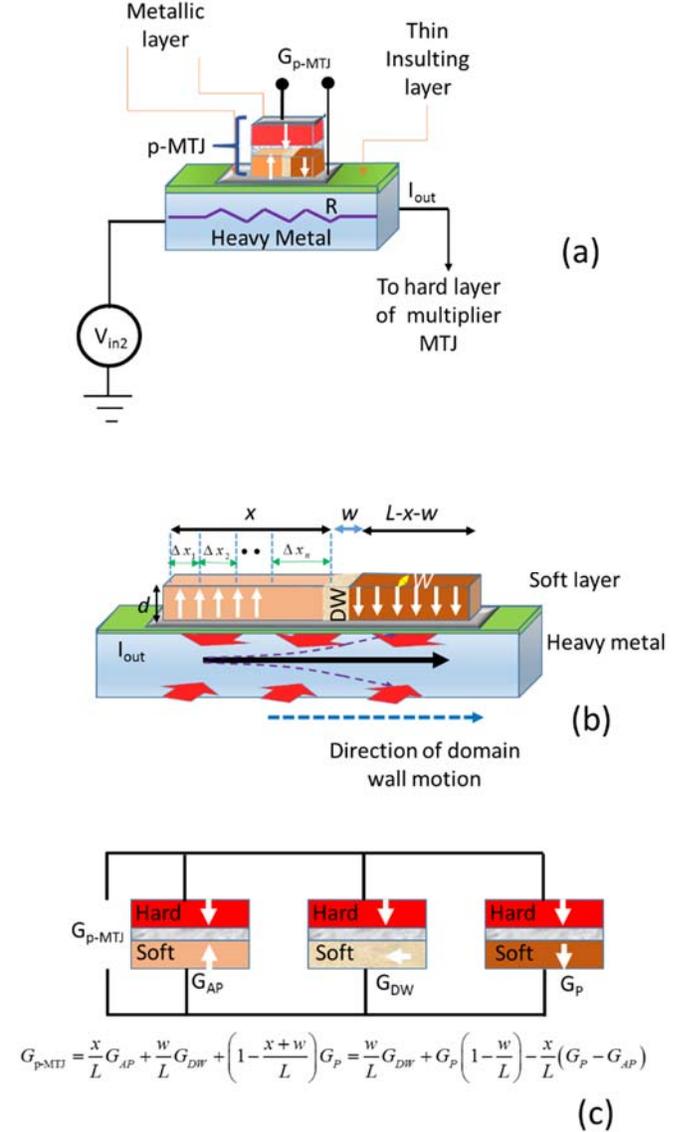

Fig. 3: (a) Schematic of the "accumulator" consisting of a p-MTJ integrated with a heavy metal strip. (b) Domain wall motion in the p-MTJ soft layer due to the flow of current through the heavy metal strip making up the resistor R in Fig. 1. (c) The conductance of the p-MTJ is the conductance of the parallel combination of three conductors associated with the anti-parallel configuration, domain wall interface, and parallel configuration.

The $m$-th current pulse will move the domain wall by an amount $\Delta x_m$ where

$$\Delta x_m = v_m \Delta t , \quad (4)$$

and $v_m$ is the domain wall velocity imparted by the $m$-th current pulse. The domain wall velocity is proportional to current density for low densities [18]. Consequently, the domain wall displacement $\Delta x_m$ will be proportional to the amplitude of the $m$-th current pulse since $\Delta t$ is fixed. We show this to be approximately true based on simulations (see Appendix A3) if the soft layer is appropriately grooved. Therefore, from Equation (4), we get

$$\Delta x_m \propto I_{out}^m \propto a_{im} \times b_{mj}. \quad (5)$$

The last equation is an important result showing that the amount by which the domain wall moves after each current pulse is proportional to the product of the two numbers $a_{im}$ and $b_{mj}$. Since $x = \sum_m \Delta x_m$, we get from Equations (3) and (5)

$$G_{\text{p-MTJ}} = \underbrace{G_{DW}\frac{w}{L} + G_P\left(1-\frac{w}{L}\right)}_{\text{constant A}} - \underbrace{\frac{(G_P - G_{AP})}{L}}_{\text{constant B}} \sum_m \Delta x_m ,$$
$$= A - B\sum_m \Delta x_m = A - B\sum_m a_{im} \times b_{mj} = A - Bc_{ij} \quad (6)$$

where $c_{ij}$ is the $(i,j)$-th element of the product matrix $[c] = [a] \times [b]$. The quantities $A$ and $B$ are constants. Finally, from Equation (6), we obtain

$$c_{ij} = \frac{A - G_{\text{p-MTJ}}}{B}. \quad (7)$$

Fig. 4 shows the composite system that constitutes the matrix multiplier to produce one element of the product matrix. In addition to the multiplier shown in Fig. 1(c) and the accumulator shown in Fig. 3(a), we use a voltage source $V_s$ proportional to $1/B$, a conductor whose conductance is equal to $A$, and another conductor whose conductance is $G_0$ where $G_0 \gg A, G_{p-MTJ}$. The current flowing through the last conductor is

$$I_{G_0} \approx \frac{-V_s}{1/G_{p-MTJ} + 1/G_0} + \frac{V_s}{1/A + 1/G_0}, \quad (8)$$
$$\approx V_s\left(A - G_{p-MTJ}\right) \propto \frac{A - G_{p-MTJ}}{B} \propto c_{ij}$$

which is proportional to the $(i, j)$-th element of the product matrix. The voltage dropped over the last conductor is proportional to this current and hence proportional to the $(i, j)$-th element of the product matrix $c_{ij}$ as well. We just have to measure this voltage after the pulse sequence has ended (i.e. the $i$-th row of the multiplier matrix has been multiplied with the $j$-th column of the multiplicand matrix) to obtain a voltage proportional to $c_{ij}$. After obtaining $c_{ij}$, the domain wall synapse is reset with a magnetic field or a reverse current pulse to make $x = 0$, and then the process is repeated to obtain the product of multiplying another row of the multiplier matrix with another column of the multiplicand matrix (which would be the next element of the product matrix). Alternately, if we do not wish to follow this process, we can dedicate one individual system in Fig. 4 for each element of the product matrix [c] in order to obtain all the elements of the product matrix in parallel. We will then need an $N \times N$ array of the units in Fig. 4 and since each unit has two devices (two MTJs), we will need $2N^2$ devices (MTJs) to multiply two $N \times N$ matrices.

### B. Energy dissipation

The energy dissipation incurred during the rotation of a nanomagnet's magnetization due to strain is very small – theoretically around 1 aJ at room temperature [15], while the energy dissipation associated with domain wall motion will be on the order of $I^2 R \Delta t$, where $I$ is the current pulse inducing the domain wall motion, $R$ is the resistance of the heavy metal strip and $\Delta t$ is the pulse width. There is some additional dissipation in the passive resistors, but they can be made arbitrarily small by choosing the bias voltages to be small. We will neglect any other dissipation due to domain wall viscosity, which would be comparatively smaller. Therefore, the energy dissipated during each MAC operation is $\sim I^2 R \Delta t$. From Fig. 1(c) we see that the current through the heavy metal strip will have a maximum value of $I_{max} = V_{in2}(\max)/(R+R_P) \approx V_{in2}(\max)/R_P$ which will have a maximum value of $\sim 50$ μA since $V_{in2}(\max) \sim 50$ mV and $R_P = 1$ kΩ. Note that this maximum current amplitude corresponds to the largest element in any matrix. There is a limit on how large an element can be and we will address that later in the Appendix.

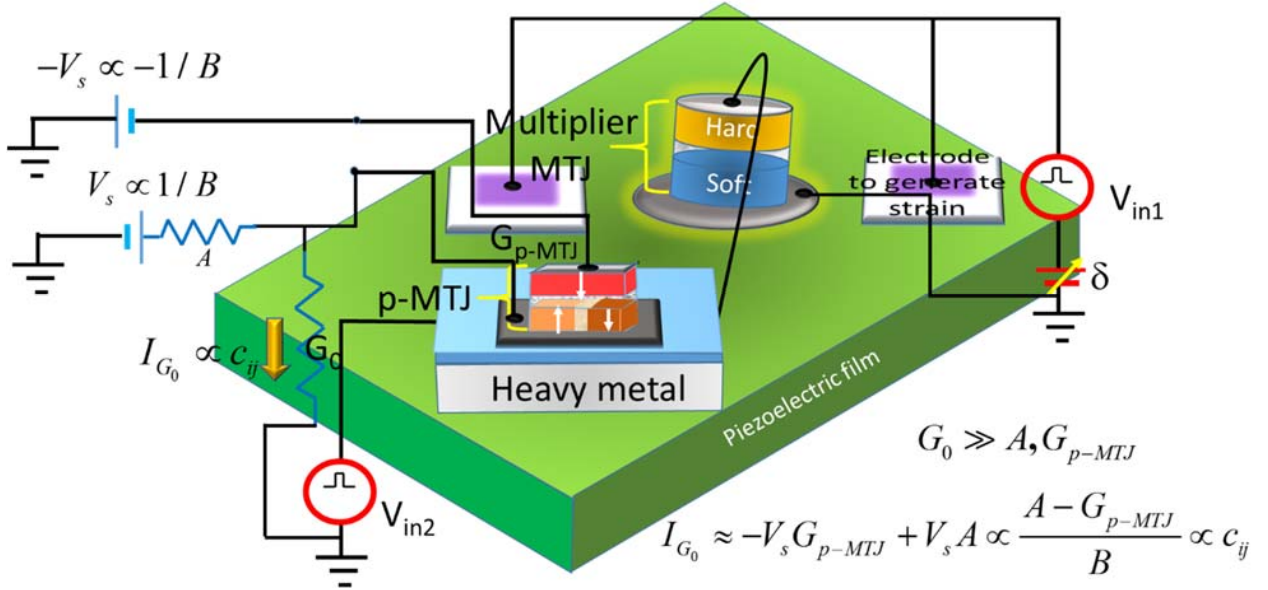

Fig. 4: The composite matrix multiplier. Vs is a battery whose voltage output is inversely proportional to $B$ and $G_0$ is a conductor whose conductance is much larger than $A$ and the conductance of the p-MTJ. The current flowing through the conductor after multiplication of one row with one column is complete is the corresponding element of the product matrix.

We will assume that the HM strip has a width of 50 nm and thickness 5 nm (cross-sectional area = 250 nm$^2$). Hence the maximum current density that will be produced in the strip is 50 μA/250 nm$^2$ = $2 \times 10^{11}$ A/m$^2$. In Appendix A3, we will show from room temperature micromagnetic simulations that the domain wall displacement at this current density is about 120 nm if we inject the current for 0.5 ns and then allow a rest period of 4.0 ns for the domain wall to stabilize.

If the maximum matrix size that one has to handle is $N_{max} \times N_{max}$, then the strip length has to be no more than $120 N_{max}$ nm to ensure that the total domain wall displacement during the generation of one element of the product matrix will not exceed the strip length. From this consideration, we calculate that the strip's maximum resistance will be $R = 48 N_{max}$ ohms, if it is made of β-Ta whose resistivity is $\sim 10^{-7}$ ohm-m. Assuming a pulse width $\Delta t = 0.5$ ns, the *maximum* energy dissipation per MAC operation is $I^2 R \Delta t = (50\ \mu A)^2 \times 48 N_{max} \times 0.5 \times 10^{-9} \approx 60 N_{max}$ aJ. In order to produce one element of a $N \times N$ product matrix, we need $N$ MAC operations and hence the maximum energy dissipated for that purpose will be $60 N_{max} \times N$ aJ. The total energy dissipated to produce all the $N \times N$ elements of the product matrix [c] will therefore be $60 N_{max} \times N^3$ aJ. If $N_{max} = 1000$, then the maximum energy dissipated to complete the matrix multiplication is 60 μJ.

*C. Comparison with an electronic crossbar matrix multiplier*

Fig. 5 shows the conventional crossbar architecture to produce one column of the product matrix, say, the *m*-th column $c_i^m\ (i=1\bullet\bullet\bullet N)$. The crossbar nodes are conductors and the current $I_i$ in the *i*-th horizontal line is given by $I_i^m = \sum_j G_{ij} V_j^m$ where $V_j^m$ is the voltage in the *j*-th vertical line and $G_{ij}$ is the conductance of the conductor at the (*i,j*)-th node. The $N \times N$ conductance matrix [G] will encode the multiplier matrix and the voltage array $V_j^m$ will encode the *m*-th column of the multiplicand matrix. Therefore, $I_i^m \equiv c_i^m\ (i=1\bullet\bullet\bullet N)$ are the elements of the *m*-th column of the product matrix. Note that in order to produce one column of the product matrix, we need $N^2$ conductors (devices). Since there are $N$ such columns in the complete product matrix, we will need $N^3$ devices to produce the entire product matrix. In comparison, the system we described requires only $2N^2$ devices, thereby offering an advantage in footprint.

The maximum energy dissipated in the crossbar array to produce the entire $N \times N$ product matrix is $\Xi N^3$, where $\Xi$ is the maximum energy dissipated in the conductor. Therefore, as long as $\Xi > 60 N_{max}$ aJ, the crossbar will be more dissipative.

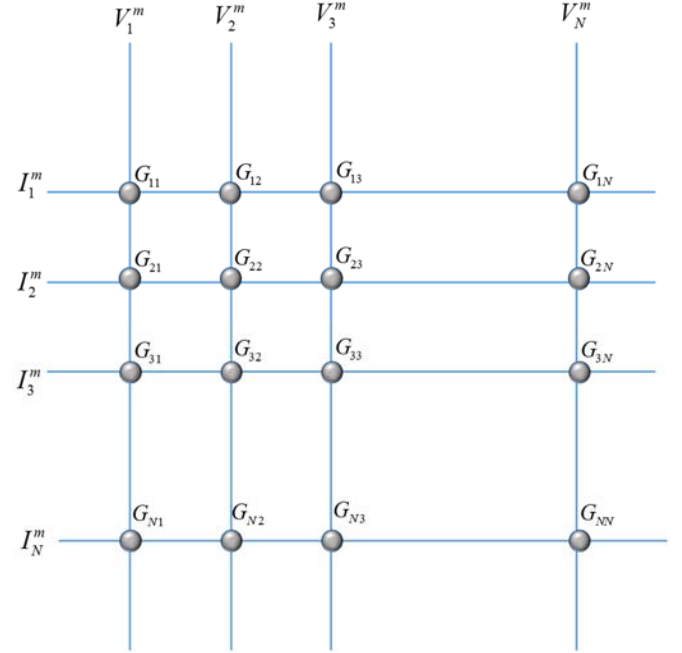

Fig. 5: An electronic crossbar architecture for matrix multiplication.

Finally, note that our proposed matrix multiplier is truly *non-volatile* since it retains information after powering off. Equation (7) shows that as soon as we turn back the constant voltage source $V_s$ after powering off, the current $I_{Go}$ flowing through the conductor $G_0$ will be proportional to an element of the product matrix. Hence, information is retained. In contrast, the crossbar architecture is volatile even if the conductance states are non-volatile, since after powering off, memory of the voltage $V_j^m$ is lost and the product matrix is not retained.

IV. CONCLUSION

We have shown how to implement a matrix multiplier with two MTJs, passive resistors and some bias sources. The energy dissipation per multiply and accumulate (MAC) operation is much smaller than what would be encountered in traditional electronic implementations. Our matrix multiplier may not be as fast as optical implementations, or even electronic implementations, but it is *non-volatile* and will retain the result of the operation (i.e. the matrix element $c_{ij}$) indefinitely after powering off. The non-volatility is a major advantage since it will allow most or all computing to be performed at the edge without the need to access the cloud. This reduces the likelihood of hacking, data loss, intrusion and eavesdropping. Cybersecurity is critical for artificial intelligence and the ability to perform all or most computing at the edge, with a non-volatile hardware accelerator, offers valuable protection against cyber threats.

The low energy dissipation per MAC operation (~$60N_{max}$ aJ) also offers protection against hardware Trojans, which are disastrous for AI and are very hard to detect. Trojans, however surreptitious, must consume some energy and hence can be detected with side channel monitoring [20], which searches for anomalies in power consumption. A *low power* matrix multiplier, which consumes very little power itself, will exacerbate power anomalies due to Trojans and facilitate Trojan detection.

## V. APPENDIX

**A.1**: We consider the elliptical soft layer of a straintronic MTJ as shown in Fig. 6. This figure shows the axis designation with the *z*-axis along the major (easy) axis of the soft layer and *y*-axis along the minor (hard) axis. We will assume that the hard layer's magnetization is along its own easy axis and is pointing along the +z-direction. In that case, the polar angle θ shown in Fig. 6 is the angle between the magnetizations of the hard and soft layers of the s-MTJ.

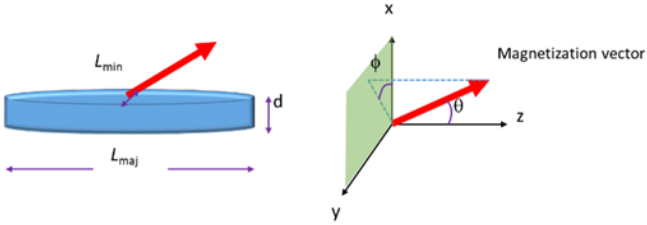

Fig. 6: The axes designation used to simulate the magneto-dynamics of the soft layer of the straintronic MTJ employed in the multiplier.

Ref. [15] showed how the stochastic Landau-Lifshitz-Gilbert equation yields the temporal evolution of the polar and azimuthal angles of the magnetization vector (θ, ϕ) in the soft layer in the presence of thermal noise and uniaxial stress. Here, we have used that recipe.

The stress σ generated in the soft layer of the s-MTJ (multiplier) by the gate voltage $V_G$ is related to the latter according to the relation $\sigma = Yd_{33}\dfrac{V_G}{T}$, where *Y* is the Young's modulus of the soft layer (Terfenol-D), $d_{33}$ is the piezoelectric coefficient of the piezoelectric film (PMN-PT), and *T* is the thickness of the film.

In the simulation, we turn on $V_G$ abruptly at time $t = 0$ and then we follow the temporal evolution of the magnetization (and hence the angle θ) until steady state is reached. Steady state is defined as the condition when θ settles to a value and fluctuates slightly around it owing to thermal noise. Because thermal noise can influence the switching trajectory (i.e. the temporal evolution of θ) from the very start, the steady state value is slightly different from run to run and hence we average over 1000 runs to find the steady-state value $\theta_{ss}$.

**A2**: ANALYTICAL DERIVATION OF THE LINEAR REGION IN THE $G_{s\text{-}MTJ}$ VERSUS $V_G$ CHARACTERISTIC

Here, we show *analytically* that in our system, the linear $G_{s\text{-}MTJ}$ versus $V_G$ relation $1/R_{s-MTJ} = 1/R_{AP} + \kappa(V_G - \delta) \Rightarrow G_{s-MTJ} = G_{AP} + \kappa(V_G - \delta)$ is obeyed in a specific region of gate voltage and derive what that region is.

The steady-state resistance of the s-MTJ is given by $R_{s-MTJ} = R_P + \dfrac{R_{AP} - R_P}{2}[1 - \cos\theta_{ss}]$, where $\theta_{ss}$ is the steady-state angle between the magnetizations of the hard and the soft layer at any given stress (or, equivalently, any given $V_G$). From ref. [15], we obtain that the magneto-static energy in the plane of the nanomagnet (i. e. when ϕ = 90°) for any magnetization orientation and at any given stress is

$$E = \left[\dfrac{\mu_0}{2}M_s^2\Omega(N_{d-yy} - N_{d-zz}) + \dfrac{3}{2}\lambda_s\sigma\Omega\right]\sin^2\theta \\ + \dfrac{\mu_0}{2}M_s^2\Omega N_{d-zz} - \dfrac{3}{2}\lambda_s\sigma\Omega + \mu_0 M_s\Omega H_d\cos\theta$$ (A1)

where $\mu_o$ is the permeability of free space, $M_s$ is the saturation magnetization, $\lambda_s$ is the saturation magnetostriction of the soft layer, σ is the stress, $N_{d-yy}$ and $N_{d-zz}$ are the demagnetization factors along the minor and major axis (they depend on the soft layer's dimensions) and Ω is the soft layer's volume. The quantity $H_d$ is the effective magnetic field in the soft layer. This field is antiparallel to the magnetization of the hard layer. The strength of this field can be tailored by engineering the material composition of the hard layer, which is usually made of a synthetic antiferromagnet, to tune the dipole interaction between the hard and soft layers. It can also be adjusted with an external in-plane magnetic field, if needed. The steady state value of the angle θ is that where the magneto-static energy in Equation (A1) is minimized.

Taking the derivative of Equation (A1) with respect to θ and setting it equal to zero, we find the angle where the energy is minimum. It corresponds to the steady state value $\theta_{ss}$. We get

$$\dfrac{\partial E}{\partial \theta} = \left[\dfrac{\mu_0}{2}M_s^2\Omega(N_{d-yy} - N_{d-zz}) + \dfrac{3}{2}\lambda_s\sigma\Omega\right]\sin(2\theta) - M_v H_d \sin\theta$$
$$= \sin\theta\left[2\left(\dfrac{\mu_0}{2}M_s^2\Omega(N_{d-yy} - N_{d-zz}) + \dfrac{3}{2}\lambda_s\sigma\Omega\right)\cos\theta - M_v H_d\right]$$
$$= 0$$ (A2)

Solving for cosθ from the above equation, we get

$$\cos\theta_{ss} = \dfrac{M_v H_d}{2\left(\dfrac{\mu_0}{2}M_s^2\Omega(N_{d-yy} - N_{d-zz}) + \dfrac{3}{2}\lambda_s\sigma\Omega\right)}$$
$$= \dfrac{M_v H_d}{\left(\mu_0 M_s^2\Omega(N_{d-yy} - N_{d-zz}) + 3\lambda_s Yd_{33}\dfrac{V_G}{T}\Omega\right)}$$ (A3)
$$= \dfrac{M_v H_d}{3\lambda_s Yd_{33}\Omega/T} \times \dfrac{1}{V_G + \mu_0 M_s^2(N_{d-yy} - N_{d-zz})T/(3\lambda_s Yd_{33})}$$
$$= \dfrac{\Gamma}{V_G - \gamma}$$

where $T$ is the thickness of the piezoelectric layer, $\Gamma = \dfrac{M_v H_d T}{3\lambda_s Y d_{33} \Omega} = \dfrac{\mu_0 M_s H_d T}{3\lambda_s Y d_{33}}$ and $\gamma = \dfrac{\mu_0 M_s^2 (N_{d-zz} - N_{d-yy}) T}{3\lambda_s Y d_{33}}$.

It is easy to verify that the second derivative $\dfrac{\partial^2 E}{\partial \theta^2}$ is positive and hence this is indeed a minimum of the energy, as opposed to a maximum.

Since a real solution of $\theta_{ss}$ is possible only if $|\cos\theta_{ss}| \leq 1$, it is obvious that $|V_G - \gamma| \geq \Gamma$. Using the values in Table I, we obtain from the above expressions that $\Gamma = 0.26$ V and $\gamma = -0.001$ V. Hence, a steady state solution for the angle between the magnetizations of the hard and soft layers (when they are not collinear) can be obtained only if $|V_G + 0.001\,\text{V}| \geq 0.26\,\text{V}$ and that is what we observe in Fig. 2(b) where the MTJ resistance begins to change only when $V_G \leq -0.261$ V.

Now, using Equation (A3), we get

$$R_{s-MTJ} = R_P + \dfrac{R_{AP} - R_P}{2}[1 - \cos\theta_{ss}]$$
$$= \dfrac{R_{AP} + R_P}{2} - \dfrac{R_{AP} - R_P}{2} \dfrac{\Gamma}{V_G - \gamma} \quad (A4)$$

and therefore

$$\dfrac{1}{R_{s-MTJ}} = \dfrac{1}{\dfrac{R_{AP}}{2}\left(1 - \dfrac{\Gamma}{V_G - \gamma}\right) + \dfrac{R_P}{2}\left(1 + \dfrac{\Gamma}{V_G - \gamma}\right)} \quad (A5)$$

When $V_G - \gamma$ is close to $-\Gamma$, we can write $\dfrac{\Gamma}{V_G - \gamma} = -1 + \varepsilon$ where $|\varepsilon| \ll 1$. Hence from Equation (A5) we obtain

$$\dfrac{1}{R_{s-MTJ}} = \dfrac{1}{R_{AP}(1-\varepsilon/2) + R_P \varepsilon/2} \approx \dfrac{1}{R_{AP}}(1+\varepsilon/2)$$
$$= \dfrac{1}{R_{AP}} + \dfrac{1}{2R_{AP}}\left(1 + \dfrac{\Gamma}{V_G - \gamma}\right) \quad (A6)$$
$$\approx \dfrac{1}{R_{AP}} - \dfrac{1}{2R_{AP}}\left(\dfrac{V_G - \gamma + \Gamma}{\Gamma}\right) \quad [\text{since } V_G - \gamma \approx -\Gamma]$$
$$= \dfrac{1}{R_{AP}} - \dfrac{1}{2R_{AP}\Gamma}(V_G - [\gamma - \Gamma])$$

Equation (A6) has the form $1/R_{s-MTJ} = 1/R_{AP} + \kappa(V_G - \delta)$ or $G_{s-MTJ} = G_{AP} + \kappa(V_G - \delta)$ where $\kappa = -\dfrac{1}{2R_{AP}\Gamma}$ and $\delta = \gamma - \Gamma$

. Thus, we have derived the existence of the linear region in the $G_{s-MTJ}$ vs. $V_G$ characteristic *analytically* and found that it exists when $V_G - \gamma$ is close to $-\Gamma$.

Since $\Gamma = 0.26$ V and $\gamma = -0.001$ V, while $R_{AP} = 2$ k$\Omega$, we find that $\kappa = -0.96$ (k$\Omega$-V)$^{-1}$ and $\delta = -0.261$ V. This value of $\delta$ shows excellent agreement with what we obtained in Fig. 2(b), but $\kappa$ is larger in magnitude by more than a factor of 2, which is still acceptable within the limits of the approximations used to derive this analytical result.

### A. Steady state value of $\theta$

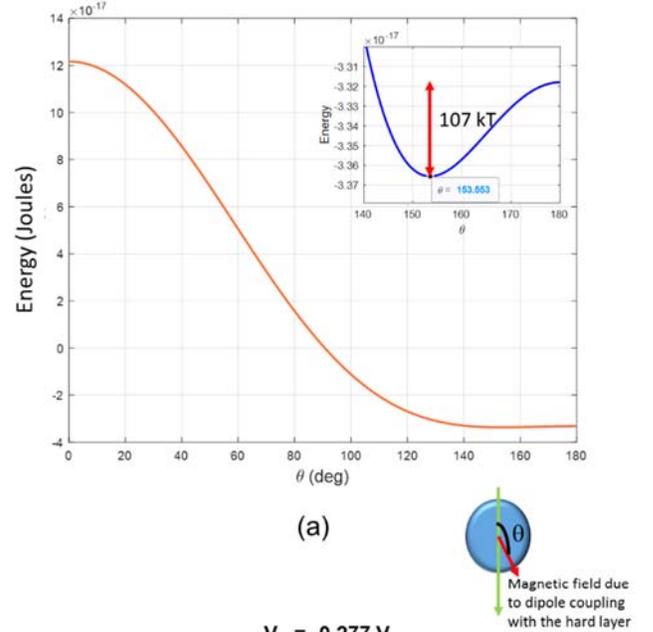

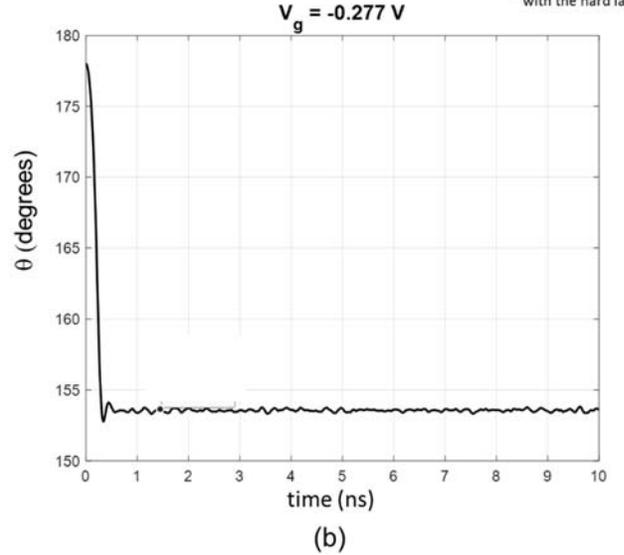

Fig. 7: (a) Potential energy in the soft layer as a function of the magnetization orientation (angle $\theta$ subtended with the major axis) when a gate voltage of -0.277 V is applied and a dipole coupling magnetic field of 1000 Oe is present along the major axis. There is a deep potential well at $\theta = 153.5^0$ which corresponds to the steady state value of $\theta$ or $\theta_{ss}$. (b) The time variation of $\theta$ after turning on the gate voltage at time $t = 0$ ns.

We show that the steady state value of $\theta$ (i.e., $\theta_{ss}$) is very stable against thermal noise. In Fig. 7(a), we plot the potential energy $E$ in Equation (A1) as a function of the angle $\theta$ at a fixed gate voltage of -0.277 V (and hence a fixed strain) assuming a magnetic field $H_d$ of 1000 Oe along the major axis. The inset shows that a deep potential well forms at $\theta = \theta_{ss} = 153.5^0$ with a depth of 107 kT at room temperature. Hence, thermal noise cannot make $\theta_{ss}$ unstable. We also show that time variation of θ in Fig. 7(b) and it becomes stable at $\theta_{ss}$.

*B. Non-binary multiplier*

The construct described here is a non-binary multiplier (meaning its elements can have integral values that are not just 0 and 1). We will of course need to know the largest integer we can have as a matrix element. That depends on how small we can make the quantization step size when we digitize the input voltage pulses $V_{in1}$ and $V_{in2}$. The minimum step size is, say, twice the thermal noise voltage appearing at any input terminal and that is $2\sqrt{kT/C_{in}}$ where $C_{in}$ is the input terminal capacitance [21]. We can reasonably assume that $C_{in} \sim 1$fF when we factor in line capacitances. This makes the minimum step size ~4 mV at room temperature. Hence the largest integer that we can encode is 50 mV/4 mV = 12. We can, of course, increase this number by using optimized design where the amplitude of the voltage pulses can exceed 50 mV. This would require decreasing κ. Here, however, we were interested in demonstrating just the basic principle and hence have not focused on design optimization. Increasing the pulse amplitude will obviously lead to more energy dissipation as well.

We can also calculate the current density through the HM strip at the minimum step size of 4 mV, which corresponds to the integer 1. The current is 4 mV/$R_P$ = 4 mV/1 kΩ = ~ 4 μA. The corresponding current density is 4 μA/250 nm² = $1.6 \times 10^{10}$ A/m², which is more than enough to induce domain wall motion in many materials [22]. In fact, the results in the next sub-section (Appendix A3) show that the domain wall displacement at this current density is about 5 nm. Hence, the smallest integer that we can have as a matrix element is 1 since the current pulse corresponding to this digit can induce sufficient domain wall motion. Thus, for this design, our integer range for any element of the $N \times N$ matrix is 1 through 12.

**A3: ROOM TEMPERATURE MICROMAGNETIC SIMULATIONS OF DOMAIN WALL MOTION IN THE SOFT LAYER OF THE ACCUMULATOR MTJ.**

It is well known that at room temperature, the domain wall motion is stochastic. After the current pulse inducing the domain wall motion subsides, the wall does not immediately stabilize, but can move forward and backward – a phenomenon sometimes referred to as domain wall creep. It is very damaging for a domain wall synapse since it will hinder the domain wall displacement from being proportional to the current amplitude, which is critical to implement the accumulator.

The solution is to make the edges of the soft layer *grooved* or notched as shown in the insets of Fig. 7 [23, 24]. They stabilize the domain wall, mitigate the effect of edge roughness in the soft layer that can trap domain walls [25], and prevent creep, but to ensure that the domain wall displacement is linearly proportional to the current amplitude (which is what we need) the pitch, depth and width of the groove will have to be chosen carefully. For this purpose, we carried out micromagnetic (MuMax3) simulations of domain well motion in the p-MTJ soft layer of dimensions 2060 nm x 50 nm x 1.5 nm and assumed a spin Hall angle of 0.2, which is reasonable when the HM is β-Ta. The soft layer of the p-MTJ is assumed to be made of CoFeB. The notch dimensions and spacing are shown in the left inset of Fig. 8. Note that the maximum matrix size we can accommodate with this choice is $N_{max} = 2060/120 = 17$.

A current pulse was injected for 0.5 ns followed by a rest period of 4.0 ns within which the domain wall position stabilized. The simulations were carried out in the presence of random thermal noise at 300 K and the mean displacements and standard deviation (error bars) of the domain wall are shown in Fig. 8 as a function of the current density injected into the HM strip. The mean and standard deviation were obtained from 100 runs of the MuMax3 (micromagnetic) simulations. The best fit straight line is shown in this plot and the points representing the mean displacements do not stray too far from this line, showing that for this choice of groove parameters, the domain wall displacement is *approximately* proportional to the current density and hence the current amplitude. This is what is needed to implement the accumulator.

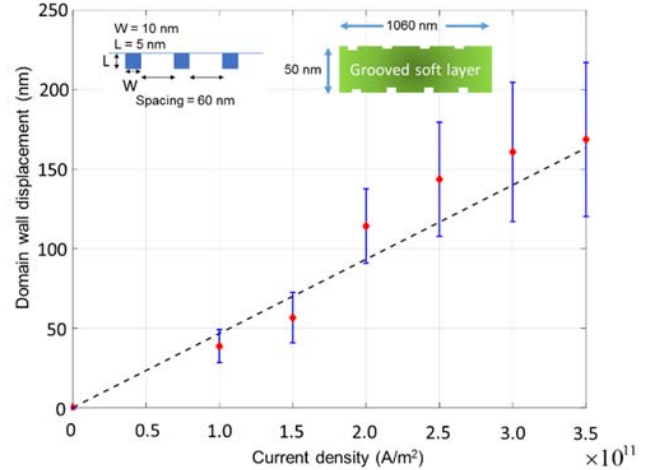

Fig. 8: Mean domain wall displacement versus current density in a grooved CoFeB soft layer. The inset shows the groove dimensions and spacing. The error bars represent the standard deviations in the domain wall displacement. Figure is not to scale. The point near the origin corresponds to the minimum current density of $1.6 \times 10^{10}$ A/m² and the domain wall displacement at this current density is ~ 5 nm.

An interesting observation is that the standard deviation in the displacements is rather large and the question naturally arises if this is a consequence of the grooved structure or thermal noise. We have examined many different groove geometries and parameters. In all cases, we saw large standard deviations and hence it is likely that choosing a different groove geometry or pattern will not reduce the standard deviation significantly. It appears that the primary culprit is thermal noise which introduces this large standard deviation.